\documentclass[prb,twocolumn,amsmath,showpacs]{revtex4}
\usepackage{graphicx}
\usepackage{color}

\begin{document}
\title{Peculiarities of the superconducting gaps and the electron-boson
interaction in TmNi$_2$B$_2$C as seen by point-contact spectroscopy}
\author{Yu. G. Naidyuk, O. E. Kvitnitskaya, L. V. Tiutrina, I. K. Yanson$\footnote{deceased}$}
\affiliation{B. Verkin Institute for Low Temperature Physics and Engineering, National Academy  of Sciences of
Ukraine,  47 Lenin Ave., 61103, Kharkiv, Ukraine}
\author{G. Behr$\footnote{deceased}$, G. Fuchs, S.-L. Drechsler, K. Nenkov,  and L. Schultz }
\affiliation{Leibniz-Institut f\"ur Festk\"orper- und
Werkstoffforschung Dresden e.V., Postfach 270116, D-01171 Dresden,
Germany}
\date{\today}

\begin{abstract}
Point-contact (PC) investigations on the title compound in the
normal and superconducting (SC) state ($T_{\rm c}\simeq10.6$\,K)
are presented. The temperature dependence of the SC
gap of TmNi$_2$B$_2$C determined from Andreev-reflection (AR) spectra using the
standard single-gap approximation (SGA) deviates from the BCS
behavior in displaying a maximum at about $T_{\rm c}/2$. A refined
analysis within the two-gap approximation provides evidence for
the presence of a second gap twice as large as the main gap (the first one),
while the latter is close to that within the SGA.
This way, TmNi$_2$B$_2$C expands the number of nickel
borocarbide superconductors which exhibit a clear multiband character.
Additionally, for the first time "reentrant" features were found
in the AR spectra for some PCs measured in a magnetic field.
The PC spectroscopy of the electron-boson interaction in TmNi$_2$B$_2$C in the normal state
reveals a pronounced phonon maximum at 9.5\,meV and a more smeared one
around 15\,meV, while at higher energies the PC spectra are almost
featureless. Additionally, the most intense peak slightly above
3\,meV observed in the PC spectra of TmNi$_2$B$_2$C is presumably
caused by crystalline-electric-field (CEF) excitations. The peak
near 1\,meV detected for some PC spectra is connected with a
modification of the CEF probably due to boron or carbon vacancies,
allowing to probe  the local stoichiometry by PC spectroscopy.



\end{abstract}

\maketitle

\section{Introduction}
Compared with the isomorphic nonmagnetic transition metal
borocarbides $R$Ni$_2$B$_2$C with $R$=Y, Lu, Sc,
La, Th, the rare-earth nickel borocarbide $R$Ni$_2$B$_2$C (here
$R$ is a rare-earth element) compounds with $R$=Dy, Ho, Er, and Tm
are of special interest due to two coexisting and competing
ordered states in these systems: antiferromagnetic (AFM) and
superconducting (SC), observed by electrical resistivity, specific
heat, magnetization, neutron, etc. measurements (see Refs.\,
\cite{Muller01,Muller02,Muller08,Gupta06} and Refs. therein). The
energy scales of these two states are of similar size, with the SC
critical temperature $T_{\rm c}$ both larger (for  $R$= Ho, Er, and
Tm) and lower (for $R$=Dy) than the AFM N\'eel temperature $T_{\rm N}$.
The SC state here is most probably of the BCS type and
mediated by the electron-phonon interaction. The magnetism in
these rare-earth compounds is provided by their localized $4f$
rare-earth ion's moments which order at
relatively high temperatures compared with their dipole-dipole
interaction due to the indirect Ruderman-Kittel-Katsuya-Yosida
(RKKY) interaction mediated by the itinerant electrons. Moreover,
the crystal electric field (CEF) splits the degenerate $4f$
rare-earth ion states. Thus, CEF effects may influence the
formation of the AFM and SC states. For a deepened understanding of
$R$Ni$_2$B$_2$C compounds spectral information both on the
electronic structure and on the bosonic collective excitations are
of importance.

Point-contact (PC) spectroscopy \cite{Naid} is a unique tool to
study simultaneously both the SC gap or order parameter and to
provide spectral data as to normal state bosonic excitations
(phonons, magnons, CEF excitations etc.) always present in these
solids. Measuring the second derivative of the $I(V)$ curves for
the PC's, information about the electron-phonon
interaction function $\alpha^2F(\omega)$ \cite{Kulik} and
analogously for other bosonic collective interactions in the normal state can be
obtained. Also the behavior of the SC gap (i.\,e. of the order
parameter in a more rigorous sense) in the SC state can be determined
from  the $dV/dI(V)$ curves using a now routine approach \cite{BTK}.

In this paper we focus on PC measurements on a rather
special member of the rare-earth nickel borocarbide family, TmNi$_2$B$_2$C, which
becomes superconducting below 11\,K. But only below $T_{\rm N}\simeq
1.5$\,K does an AFM ordering appear where transversely polarized
spin-density waves with an incommensurate modulation of the
magnetic moments sets in. \cite{Muller01,Muller02,Muller08,Gupta06} Thus TmNi$_2$B$_2$C
exhibits the largest difference between $T_{\rm c}$ and $T_{\rm N}$ in
the whole $R$Ni$_2$B$_2$C series, which points to a
relatively weak magnetism which cannot destroy
the superconductivity on large Fermi surface sheets (FSS's), but only weakens it,
although sizably, strongly dependent on the purity of the single crystals
and as well as on the field direction.\cite{Muller01,Muller02,Muller08,Gupta06}
The superconductivity is nevertheless affected by the special
{\it incommensurate } magnetic state and its fluctuation above
$T_{\rm N}\simeq 1.5$\,K, at least for somehow
disordered samples (probably with respect to C--B local disorder).
In this context the {\it short} periodicity of the incommensurate
antiferromagnetic structure, $\approx$ 25\,\AA, as
compared with the superconducting coherence length, $\xi \approx$ 125\,\AA,
is of interest \cite{gamel} and a further argument given there
that magnetism and superconductivity affect each other.
In the present paper we do not consider the very coexistence of magnetism and
superconductivity because this is already outside the range of our experimental
setup ($T > 1.6$\,K).
For the temperature
range above $T_{\mbox{\tiny N}}$ and at least for low external magnetic fields
one may also expect to detect remnants of the multiple-gap
features well-known for non-magnetic borocarbides.
\cite{shulga,BobrovLu,Bashlakov,raychaudhuri} A clear hint for
such a multiple-gap behavior is given by experimentally observed positive
curvature pronounced here for the upper critical field applied
$\parallel$ to the $ab$-plane. Due to the large magnetization for
$H \parallel$ to the $c$-axis, $H_{c2}$ is most strongly suppressed
and its positive curvature as well. \cite{brison}

To the best of our knowledge, no PC data concerning
TmNi$_2$B$_2$C, are available, except our recent preliminary
results. \cite{Naid07,Naid08} On the contrary, there are several papers
devoted to investigation of other members of the nickel
borocarbide family (see Refs. \cite{naidyuk7,Naid07,Naid08} and Refs.
therein), where the SC gap as well as the electron-phonon spectral
function were studied. The peculiar non-monotonic temperature
dependence of the upper critical field $H_{c2}(T)\parallel $ to
the $c$-axis observed frequently \cite{cho} has been attributed to
Pauli paramagnetic effects \cite{debeer-schmitt}. All this makes
investigations of the missing TmNi$_2$B$_2$C by PC spectroscopy
desirable in order to get a complete picture. Here we restrict
ourselves to a study of the paramagnetic phase above $T_{\rm N}\simeq 1.5$\,K.

\section{Experimental details}
TmNi$_2$B$_2$C single crystals were grown by the float-zone
technique with optical heating. \cite{Souptel} For homogenization
and in order to minimize the amount of possible disorder on the
boron and carbon sites, the crystals were wrapped into tantalum
foil and subjected to a heat treatment at 1000$^\circ$~C for 72~h
and subsequently at 500$^\circ$C for 72~h under purified argon
atmosphere. The residual resistance ratio [RRR =$\rho$(300\,K)/$\rho$(11\,K)]
of the investigated crystal significantly increased by this heat treatment:
from RRR=14 before heating to RRR=28 after it. The high perfection of the single
crystals is confirmed by the small width (about 0.45 $\deg$) of
neutron diffraction peaks obtained for (103) Bragg reflection. The
investigated single crystal had a resistivity of $\rho$ (11\,K) = 1.7 $\mu \Omega$\,cm
in the normal state. Its SC transition
temperature amounts $T_{\rm c}$ = 10.6\,K, determined from the midpoint
value of the SC transition.

The PC spectra have been measured mainly along the $c$
direction by a standard "needle-anvil" method, \cite{Naid} touching a
cleaved TmNi$_2$B$_2$C surface with a sharpened thin Cu wire. A
number of PC spectra has been measured perpendicular to the $c$
direction, but we have found no qualitative difference between
them and the PC spectra along the $c$ direction. The reason might
be that the cleaved surface was quite irregular and rough,
therefore the crystallographic direction in PC is
not well defined and the PC spectra are expected to
be averaged over a large angle range.


The differential resistance $dV/dI(V)$ and the second derivatives
$d^2V/dI^2(V)$ were recorded through sweeping the $dc$ current $I$
on which a small $ac$ current $i$ was superimposed. The alternating
voltage $V_1\propto dV/dI(V)$ as well as its second harmonic
$V_2\propto d^2V/dI^2(V)$ have been recorded using the standard
phase sensitive lock-in technique. It is well known that the first
derivative $dV/dI(V)=R(V)$ of the $I(V)$ curve between
a normal metal and a superconductor, the so-called
Andreev-reflection (AR) spectrum, reflects the SC gap.
\cite{BTK} The second derivative of the
$I(V)$ curve of the type $R^{-1}dR/dV=R^{-2}d^2V/dI^2$ is directly
proportional to the electron-boson spectral
function according to the PC spectroscopy theory (see Ref. \cite{Naid},
chapter 3 and Ref. \cite{Kulik}). This derivative can be readily
obtained: simply by dividing the measured $V_2$ by $V_1^2$, i.e.,
$R^{-1}dR/dV\propto V_2/V_1^2$.

\section{PC spectroscopy of the SC energy gap}

\begin{figure}[t]
\vspace{3cm}
\begin{center}
\includegraphics[width=8.5cm,angle=0]{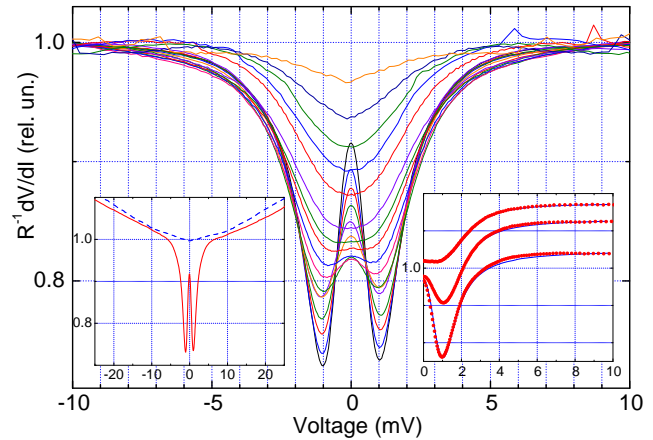}
\vspace{-3cm}
\end{center}
\caption[]{(Color online) Normalized to the
normal state and symmetrized $dV/dI$ curves of a TmNi$_2$B$_2$C--Cu
contact ($R=5.7\,\Omega$) established approximately in the $c$
direction at temperatures from the bottom to the top: 1.6\,K,
2\,K, 2.5\,K ... and further with 0.5\,K step until 9.5\,K (upper
curve). Left inset: Example of a raw $dV/dI$ curve at 1.6\,K (solid curve)
and 10\,K (dashed curve) extended to the higher biases. Right inset: Some selected
normalized and symmetrized $dV/dI$ for the same PC (points) at
$T$=1.6, 3.5 and 6\,K (from the bottom to the top) along with
calculated theoretical (according to the theory \cite{BTK}) curves
(thin lines).} \label{dvdi}
\end{figure}

To obtain a spectroscopic insight into the SC energy gap as well as
into the spectral function of the electron-boson interaction, the
contact size $d$ should be smaller than the inelastic electron
mean-free path. To avoid a variation of the SC gap on the scale of
the PC size, also the contact diameter $d$ should be less than
the coherence length $\xi$, which is in TmNi$_2$B$_2$C between
\,11 and 15\,nm. \cite{cho,Eskildsen,gamel}
The elastic electronic mean-free path in TmNi$_2$B$_2$C is
estimated as about $l \simeq 20$ nm, using typical
for the nickel borocarbides $\rho l \simeq 3.6\times
10^{-12}\Omega$\,cm$^2$ (see Ref. \cite{Bhat}) and our residual resistivity
$\rho_0\simeq 1.7\mu\Omega$\,cm. According to Wexler's formula
\cite{Wexler} $R_{\rm {PC}}\simeq 16\rho l/3 \pi d^2+ \rho/2d$,
(here we neglected contribution to the second
term from the resistivity of a normal metal (needle) and suppose
geometrically equal parts occupied by each metal in the case of
heterocontact), the PC diameter $d$ can be estimated from
their resistance $R_{\rm {PC}}$ and $\rho l$.
Using a typical $R_{\rm {PC}}$ between 1\,$\Omega$ and
10\,$\Omega$, one estimates a $d$ value between 30\,nm and 10\,nm.
Thus, the necessary condition of the smallness of the PC size
compared to $l$ and  $\xi$ may be fulfilled for the investigated PCs. Even
more importantly from the experimental or practical point of view
is the observation of characteristic AR features
in the $dV/dI(V)$ spectra in the SC state along
with the phonon structure in $d^2V/dI^2(V)$ in the normal state,
which can be used as criteria for a spectroscopic
(ballistic or diffusive \cite{Naid}) regime in PC.
\begin{figure}[t]
\vspace{0cm}
\begin{center}
\includegraphics[width=8cm,angle=0]{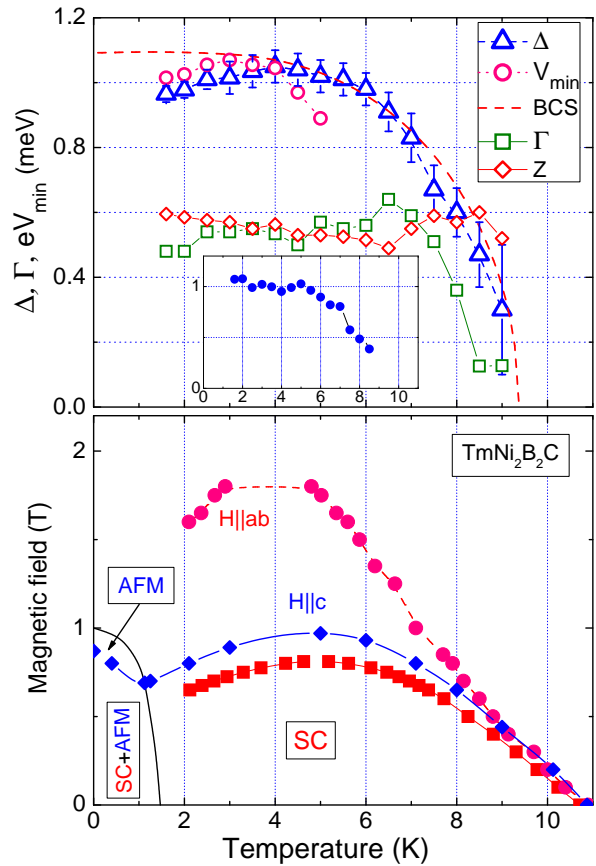}
\vspace{-0.5cm}
\end{center}
\vspace{0cm}\caption[] {(Color online) Upper
panel: Temperature dependencies of the fitting parameters (within
the SGA) for the PC spectra from Fig.\,\ref{dvdi}: the SC gap
$\Delta$ (triangles), the broadening parameter $\Gamma$ (squares),
the barrier parameter $Z$ (diamonds) as well as the scaling factor
$S$ in the inset. The minimum position in $dV/dI$ is shown by
circles. Bottom panel: Phase diagram for both superconductivity
and AFM spin-density wave order of TmNi$_2$B$_2$C using data from
\cite{gamel} along with our critical magnetic field data along the
c-axis (squares) and ab-plane (circles).
}
\label{DGZ}
\end{figure}

We succeeded in obtaining high-quality $dV/dI$ PC characteristics as
shown in Fig.\,\ref{dvdi}, which demonstrates clear AR features --
deep minima around $V\simeq \pm\Delta/$e at $T\ll T_{\rm c}$,
here $\Delta $ is the SC gap, and the absence of
any other irregularities like spikes, humps, etc. For most PCs, the
structure (minimum) which characterizes the SC
state disappears at about 10\,K, slightly below
the mentioned bulk $T_{\rm c}\simeq 10.6\pm0.05$\,K.
The experimental $dV/dI$ curves were fitted according to the well-known
modified BTK theory (see Ref. \cite{Naid}, section 3.7) within the single gap approximation (SGA).
In the inset of Fig.\,\ref{dvdi}, as an example,
the calculated $dV$/$dI$ along with the measured ones are shown at
several temperatures. From the fitting of experimental $dV/dI$
at different temperatures, the temperature dependencies of the SC
gap $\Delta$, the so-called Dynes or broadening parameter $\Gamma$, \cite{Dynes78}
the barrier strength $Z$, and the scaling factor $S$ have been determined
(see Fig.\,\ref{DGZ} and Appendix).

As shown in Fig.\,\ref{DGZ}, $\Gamma$ and $Z$ are nearly constant,
excluding the region close to $T_{\rm c}$. However, in this region
above 6\,K the double minimum structure in $dV/dI$ is washed out
and the self-action computer procedure gives more
room for the fitting parameters by approaching $T_{\rm c}$. The
most interesting feature is the temperature dependence of the gap
$\Delta(T)$. As shown in Fig.\,\ref{DGZ}, $\Delta(T)$ deviates
from the BCS-type behavior, reaching a maximum not
at $T\to0$, but at about $T\simeq 5$\,K $\simeq T_{\rm c}/2$. The
majority of investigated PCs reveals a similar behavior for the
extracted $\Delta$. Moreover, the minima positions in $dV/dI$
which at low temperatures should be close to the  $\Delta$ value
also display  maximum (see Fig.\,\ref{DGZ}) directly supporting
the obtained specific $\Delta (T)$ behavior in this temperature
region. The upper critical field $H_{c2}(T)$ measured for fields  along the
$c$ direction and within the $ab$ plane behaves similarly (see
Fig.\,\ref{DGZ}), that is, it has a maximum. This broad maximum
is closely related to the reentrant behavior found in the temperature dependence
of both the bulk resistivity (see inset of Fig.\,\ref{field}) and of the magnetic
susceptibility of TmNi$_2$B$_2$C in magnetic field.
Obviously, the recovery of the normal state at low temperature in the presence of a
sufficiently high applied magnetic field is responsible for the anomalous decrease of
$H_{c2}(T)$ at low temperatures. This $H_{c2}(T)$ anomaly might support
our finding. In fact, since in the clean limit one has in the orbital
approximation (ignoring a possible Pauli limiting behavior mentioned
above \cite{debeer-schmitt}) $H_{c2}(T) \propto \Delta^2(T)$,
the anomalous decrease of the upper critical field is more pronounced in the gap itself.


\begin{figure} [tbp]
\vspace{3cm}
\begin{center}
\includegraphics[width=8cm,angle=0]{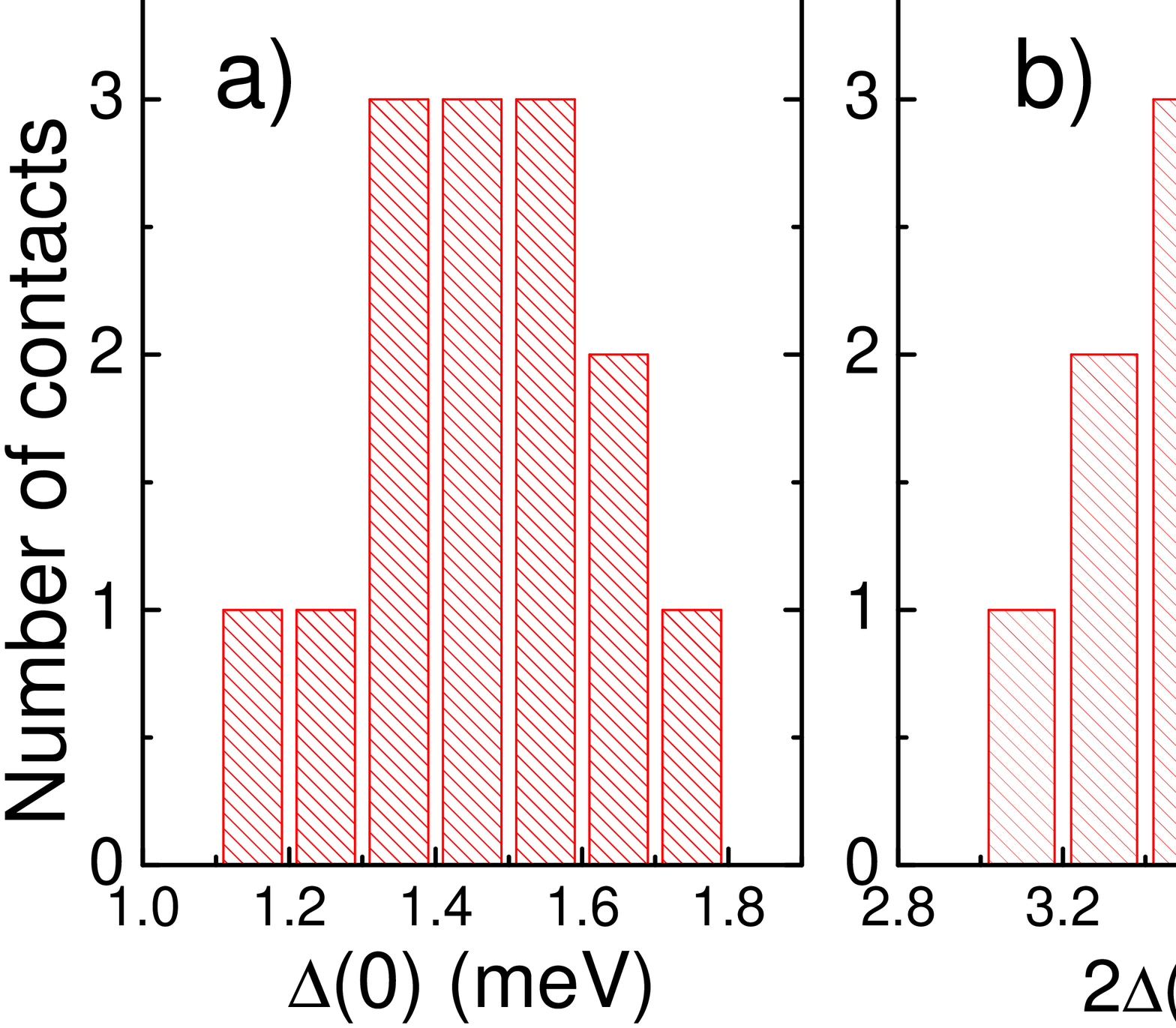}
\end{center}
\vspace{-3cm} \caption[] {(Color online) Distribution of the
single SC gap $\Delta(0)$ extrapolated by the BCS curve to $T$=0\,K
(left panel) and the characteristic ratio $2\Delta(0) / $k$_{{\rm B}} T_{\rm c}$
(right panel) for the measured contacts.
Notice the different shapes of both histograms
reflect the different $T_{\rm c}$ values of various probed PCs.}
\label{hist}
\end{figure}

The gap distribution for the selected 14 contacts
with pronounced AR minima  along with the
characteristic ratio $2\Delta/$k$_{\rm B} T_{\rm c}$ are shown in
Fig.\,\ref{hist}. The SC gap value $\Delta_0$ (in the SGA
extrapolated by the BCS dependence to $T$=0) results in a somewhat
wide range $\Delta_0=1.1-1.7$\,meV, which reveals a characteristic
ratio $2\Delta / $k$_{\rm B} T_{\rm c}$  between 3 and 4.2,
\footnote{We measured systematically $dV/dI$ with AR minima position close to 1\,mV (see,
e.\,g., Fig.\,\ref{dvdi}), corresponding to $2\Delta_0/$k$_{\rm B}
T_{\rm c}\simeq$ 3, which may be connected with both anisotropy
and/or a multiband scenario. See also the further section concerning
two-gap calculations.} here
$T_{\rm c}$ is the temperature where the AR features for the
corresponding contact vanish, in other words, the local $T_{\rm c}$. 
It varies between 9 and 10\,K. Then the mean
$2\Delta_0/$k$_{\rm B} T_{\rm c}$ exceeds the BCS weak coupling
value 3.52, which points quite generally to a strongly coupled SC
state. Alternatively, it might also reflect simply the local gap
anisotropy on a given FSS. In order to explain the nearly
isotropic $H_{c2}$ in DyNi$_2$B$_2$C \cite{peng,tomy}
(see, e.g., Fig.\,6 in Ref.\,\cite{jensen}), a somewhat
anisotropic electron-phonon interaction and also a corresponding
gap anisotropy must be assumed to compensate the {\it different}
anisotropy of the Fermi velocities of the calculated cushion FSS,
which would also affect $H_{c2}$. A similar enhanced gap ratio is also
characteristic  for other $R$Ni$_2$B$_2$C compounds
\cite{Naid07,Naid08} measured by AR spectroscopy. The gap
distribution in TmNi$_2$B$_2$C might be connected both with the
anisotropy and the multiband superconductivity, as suggested
by the analysis of the PC data for the nonmagnetic LuNi$_2$B$_2$C,
\cite{BobrovLu,Kvit10}  YNi$_2$B$_2$C, \cite{Bashlakov} and the AFM
ErNi$_2$B$_2$C \cite{Bobrov08} superconductors, or for the well-known
two-gap superconductor MgB$_2$. \cite{MgB2} On the other hand,
existing tunneling spectroscopy data for TmNi$_2$B$_2$C
\cite{Suderov01} read: $\Delta_0=1.45$\,meV and
$2\Delta_0/$k$_{\rm B} T_{\rm c}\simeq$ 3.1 with error bars of
$\pm$ 15\%. Additionally, they did not show a remarkable
anisotropy and exhibit a BCS-like weak-coupling temperature
dependence of the gap in the whole temperature range, even in the
AFM region below 1.2\,K.

\begin{figure}[tbp]
\vspace{3cm}
\begin{center}
\includegraphics[width=8.5cm,angle=0]{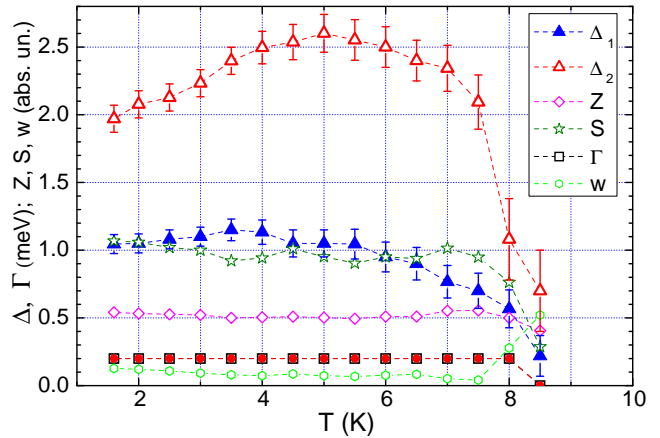}
\vspace{-3cm}
\end{center}
\vspace{0cm}\caption[] {(Color online) Temperature dependencies of
the fitting parameters in the two-gap approximation (TGA): small
(closed triangles) and large (open triangles) the SC gaps
$\Delta_{1,2}$, the broadening parameters
$\Gamma_1=\Gamma_2=0.2$ (squares), the barrier
parameter $Z$ (diamonds), the scaling factor $S$ (stars), and the
contribution (weight factor) $w$ to the d$V/$d$I$ curve of the
large gap (open circles).
Error bars for the gaps show the difference
between the presented gap values and the gaps calculated supposing
$\Gamma_2\simeq 2\Gamma_1$ (see also footnote Ref. \#54).}
\label{DGZ2}
\end{figure}

Analyzing in more details our data and the SGA fitting parameters,
it turned out that: 1) usually a quite large $\Gamma \geq
\Delta/2$ is needed to fit the $dV/dI$ curves, 2) the scaling
factor S (see Appendix) varies considerably (see, e.\,g.,
Fig.\,\ref{DGZ}), 3) there is a wide distribution of the gaps.
The large $\Gamma$ might point to a spatial (or
directional) gap distribution due to anisotropy or the multiband
character of the SC state, a picture which is now
widely accepted in the borocarbide community.
Therefore, supposing a multiband SC state we have also carried out
a two-gap fit.
{\footnote{The two-gap(band) approach is supported also
by a recent three-dimensional study of the Fermi surface of LuNi$_{2}$B$_{2}$C
\cite{Dugdale09}, where the contributions to the total density-of-states (DOS)
at the Fermi energy from three bands equal to 0.24\%, 22.64\% and 77.1\%, respectively,
have been found. That is, basically only two bands contribute to the total DOS.}}
In general, this fit is similar to the one-gap fit shown in
Fig.\,\ref{dvdi}(right inset), only the two-gap fit is a bit better in the 4--6\,mV range
for the $dV/dI$ curves at the lowest temperatures.
The results are shown in Fig.\,\ref{DGZ2}. First of all we note
 that 1) $\Gamma$ is constant and more than two times
smaller than that of our SGA fit
{\footnote{At first glance it seems naturally to suppose a two times
larger $\Gamma_2$ for the twice as large second gap. However, in this case the
contribution of the second gap (or weight factor) $w$ strongly varies at low temperature
(about 4 times): it decreases from 0.32 to 0.09 with temperature rising
from 1.6\,K to 4\,K. In our opinion there is no physical
reason for such behavior.}}
and 2) the scaling factor is close to 1 and varies moderately, except the
region close to $T_{\rm c}$, where the fit gives more freedom for
fitting parameters due to a temperature broadening and smearing of
a double-minimum structure of $dV/dI$. These two observations
suggest that the obtained results of our
two-gap fit are physically reasonable. As it is seen
from Fig.\,\ref{DGZ2}, the small gap exhibits a more or less
similar behavior as the gap obtained in the SGA fit, while the
second gap is two times larger, but its contribution to the
spectra is about 10\%, once again excluding the region
close to $T_{\rm c}$. Thus we found plausible
arguments for the presence of a second SC gap in TmNi$_2$B$_2$C. To clarify its
nature, detailed directional PC investigations are
desirable.

In the end of this section we would like to mention that except YbNi$_2$B$_2$C
with heavy-fermion like properties, at first glance
the electronic structure as predicted by the first electronic structure
calculations based on the local density approximation (LDA) is
hardly affected by the rare-earth component directly. \cite{Dugdale09}
In particular, all these compounds exhibit a special FSS,
called in the case of $R$=Lu, Y, Ho and Dy
as "cushion"\cite{bergk} or "pillow"\cite{drechsler4} which
is formed only by Ni 3$d$ states, i.\,e., without an admixture of
rare-earth 5$d$ derived states. The latter mediate the RKKY
exchange coupling of the magnetic moments of the 4$f$ electrons to
some of the conduction electrons whose superconductivity is then
strongly affected by the presence of the localized rare-earth
magnetic moments. On the contrary, the superconductivity on the
"cushion" FSS is almost perfectly protected against local magnetic
fields caused by the exchange interaction between its conduction
electrons and the 4$f$-moments.
However, recent band structure calculations for TmNi$_2$B$_2$C show explicitely
that some FSS has somewhat changed their shape. In particular,
the "cushion" has not been detected. \cite{rosnerprivate}
By symmetry an analogous Ni-$3d$ derived FSS free of an admixture
with Tm $5d$ states should nevertheless exist. Since its exact shape
has not been resolved yet, we denote it, for the sake of simplicity,
as "pseudo-cushion"(pc)-FSS. On these grounds a similar local
SC gap of $\approx $ 1\,meV can be expected for that pc-FSS in
TmNi$_2$B$_2$C, too. The PC data for the smaller gap reported above
(which is less sensitive to the vicinity of magnetism)
provide strong support for such an assignment in the title compound, too.
The larger second gap reported here, too, with a stronger sensitivity
to the vicinity of magnetism corresponds to the major FSS in nonmagnetic
borocarbides, which in the present case is somewhat affected by the
fluctuations of this AFM state occurring below 1.6\,K.

\subsection{"Reentrant" features for $dV/dI$ characteristics in an external magnetic field}

\begin{figure} [tbp]
\vspace{0cm}
\begin{center}
\includegraphics[width=8.5cm,angle=0]{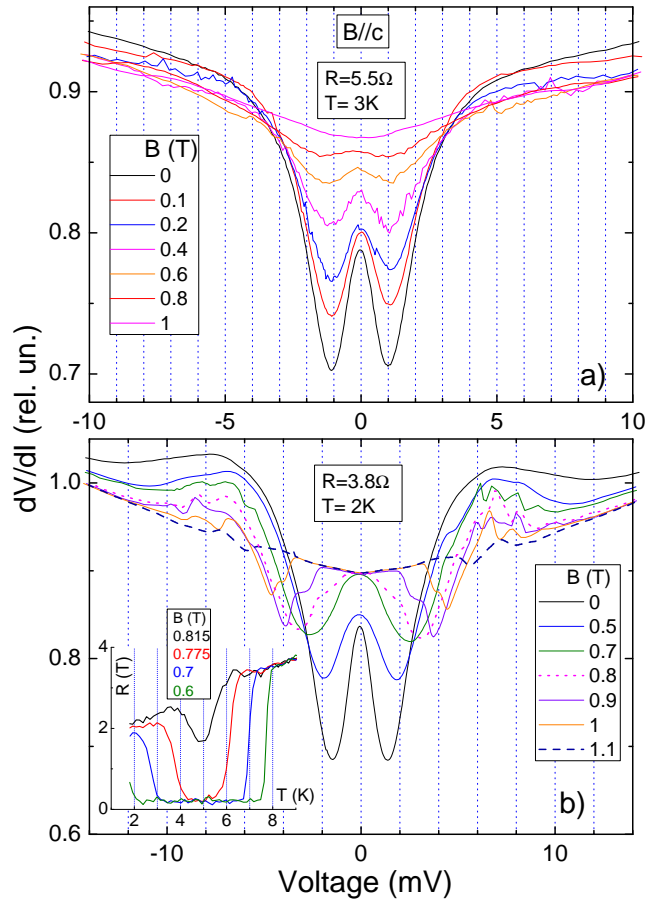}
\end{center}
\vspace{0cm} \caption[] {(Color online) a) Raw $dV/dI$ curves of
contact from Fig.\,\ref{dvdi} (after the
temperature measurements) measured at $T$=3\,K in magnetic field.
b) Raw $dV/dI$ curves of another TmNi$_2$B$_2$C--Cu contact
measured at $T$=2\,K in a magnetic field. Note the transformation
of the zero-bias maximum into a plateau-like structure above
0.7\,T. Inset: reentrant features in the resistance of a
TmNi$_2$B$_2$C single crystal in an external magnetic field
$\textbf{B}\parallel\textbf{c}$.} \label{field}
\end{figure}

Let us turn to the measurements of $dV/dI$ in an external magnetic
field. By studying the obtained curves one can distinguish two types
of $dV/dI$ behavior. One of them is shown in the upper panel of
Fig.\,\ref{field}. Here the external magnetic field does not
substantially change the position of the  $dV/dI$ minima (reflecting the
SC gap) up to fields of about 1\,T. The magnetic field reduces
the intensity of the whole $dV/dI$ minimum and its influence looks
like a smearing of the AR minima. A qualitatively similar behavior
is observed for most of PCs based on the title compound and on
many other superconductors, including nickel borocarbides
\cite{Bashlakov,raychaudhuri,naidyuk7,XLu} as well as simple SC
metals, e.\,g., Nb \cite{Naid96nb}. In the meantime the
$dV/dI$ for some studied PCs have demonstrated a quite different
behavior, \footnote{The PC spectrum for this contact is shown in Fig.\,\ref{d2vdi2}.
It demonstrates clear CEF and phonon maxima, therefore it is unlikely
that specific $dV/dI$ behavior is connected with a regime in PC that is not spectral (e.g., thermal). In our case humps in $dV/dI$ around 6\,mV are likely connected with
critical current or critical (Oersted) field effects.}
similar to that shown in Fig.\,\ref{field}(b). Here, the
double-minimum structure undergoes significant modifications in a
magnetic field. Below 0.6\,T the d$V$/d$I$ curves behave like in
the previous case, but at higher field they transform into another
structure with a plateau-like maximum at zero bias, so that the
zero-bias resistance reaches its normal-state value
at 0.7\,T and then this normal-state region spreads out to higher voltages  by
further field increase until all $dV/dI$ curves transform into a
parabolic-like normal-state behavior.

This structure of $dV/dI$ in Fig.\,\ref{field}(b) looks similar to
reentrant behavior found in the temperature
dependence of the bulk resistivity of TmNi$_2$B$_2$C in the external magnetic field
\footnote{We note, that the reentrant behavior in the magnetic
field is not observed for all single crystals
(according to unpublished data in Ref.\, \cite{Lamprecht})
and does not occur for very clean samples.
Moreover, the critical field can differ by a factor up to 2 even
for annealed samples, which can be connected with difficulties to
control the stoichiometry and the appropriate site position for
boron and carbon.},
where the normal state is restored at 2\,K in
a field above 0.7\,T (see inset in Fig.\,\ref{field}b).
This behavior of the bulk resistivity, i.\,e., the recovering of the normal
state at low temperature under a magnetic field, may be connected with some
kind of magnetic ordering induced by the external magnetic field.
The temperature increase (e.\,g., above 2\,K at 7\,T, see inset in Fig.\,\ref{field}b)
destroys this state and recovers the SC state. We assume that in our case the applied
voltage plays a role similar to the temperature, namely, a relaxation of strongly
nonequilibrium electron distribution function in a PC under an applied voltage
\cite{Kulik} can produce nonequilibrium phonons with an effective temperature $\leq$eV/4 \cite{Kulik1}.
Therefore in our case applied voltage increase recovers the SC state (SC features in d$V$/d$I$) in the field above 7\,T at 2\,K.

This qualitatively different behavior observed in
both types of $dV/dI$ characteristics is consistent with the observation that reentrant behavior of $\rho (T)$ in a magnetic field is not found for all
TmNi$_2$B$_2$C single crystals. \cite{Lamprecht}


\begin{figure} [tbp]
\vspace{3cm}
\begin{center}
\includegraphics[width=8.5cm,angle=0]{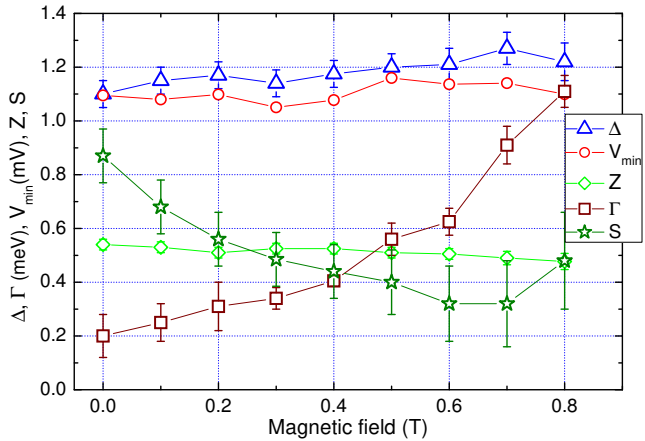}
\end{center}
\vspace{-3cm} \caption[] {(Color online) Magnetic field
dependencies of the fitting parameters within the SGA
taken at $T$=3\,K: the SC gap
$\Delta$ (triangles), the broadening parameter $\Gamma$ (squares),
the barrier parameter $Z$ (diamonds), the scaling factor $S$
(stars), as well as the minimum position in $dV/dI$ (circles)
extracted from the PC spectra in Fig.\,\ref{field}(a).}
\label{deltaf}
\end{figure}
We have also recovered the magnetic field dependence of the SC gap
and of the other fitting parameters again in the SGA. The results
are shown in Fig.\,\ref{deltaf} using the spectra from
Fig.\,\ref{field}(a). It turned out that the effective single gap is
almost magnetic field {\it independent}, strictly speaking, it
exhibits even a slight increase. This observation is in accord
with the minima position in $dV/dI$. However, this unusual,
absolutely counterintuitive and striking behavior (never observed
for any superconductor to the best of our knowledge)
might be ascribed to an unresolved anisotropic multigap scenario. In fact,
provided the PC reflects also some $(ab)$-geometry, a gap
contribution from non-cushion FSS's with a much higher upper
critical field $H_{c2}$ than the depicted highest field of
1\,T might readily resolve the puzzle.

$\Gamma$ steeply increases in an external magnetic field as it has been
found for many other type-II superconductors, \cite{Naid,Naid96nb}
while in type-I superconductors $\Gamma$ is almost field
independent. \cite{Naid96} The increase of $\Gamma$ is usually
connected with a shortening of the bosonic (Cooper
pairs) life-time due to the pair-breaking effect
of an external magnetic field. \cite{Tinkham} The
scaling parameter $S$ decreases with the magnetic field. This can
be understood as a shrinkage of the SC volume in the PC region,
probably due to vortex penetration. However, the determination of
the SC gap and other parameters from AR spectra in a magnetic
field using the standard BTK approach with the introduction of the
SC density of states smearing due to life-time effects
\cite{Dynes78} only is, in principle, a simplified approach. As mentioned in Ref.
\cite{Miyoshi}, the SC density of states in the mixed state varies
in space, i.\,e., becomes {\it nonuniform}. This special situation
requires an appropriate theoretical description (not available at present)
for the density of states to compute realistic AR spectra. Moreover,
the local SC density of states in PC will depend on the position of the pinned vortex with
respect to the PC core. Therefore the obtained effective
parameters of the homogeneous model and their
magnetic field dependence extracted from AR spectra measured in
an external magnetic field above the lower critical field $H_{c1}$
should be interpreted with some caution.

\section{PC spectroscopy of bosonic excitations in the normal state}

\begin{figure} [tbp]
\vspace{3cm}
\begin{center}
\includegraphics[width=8.5cm,angle=0]{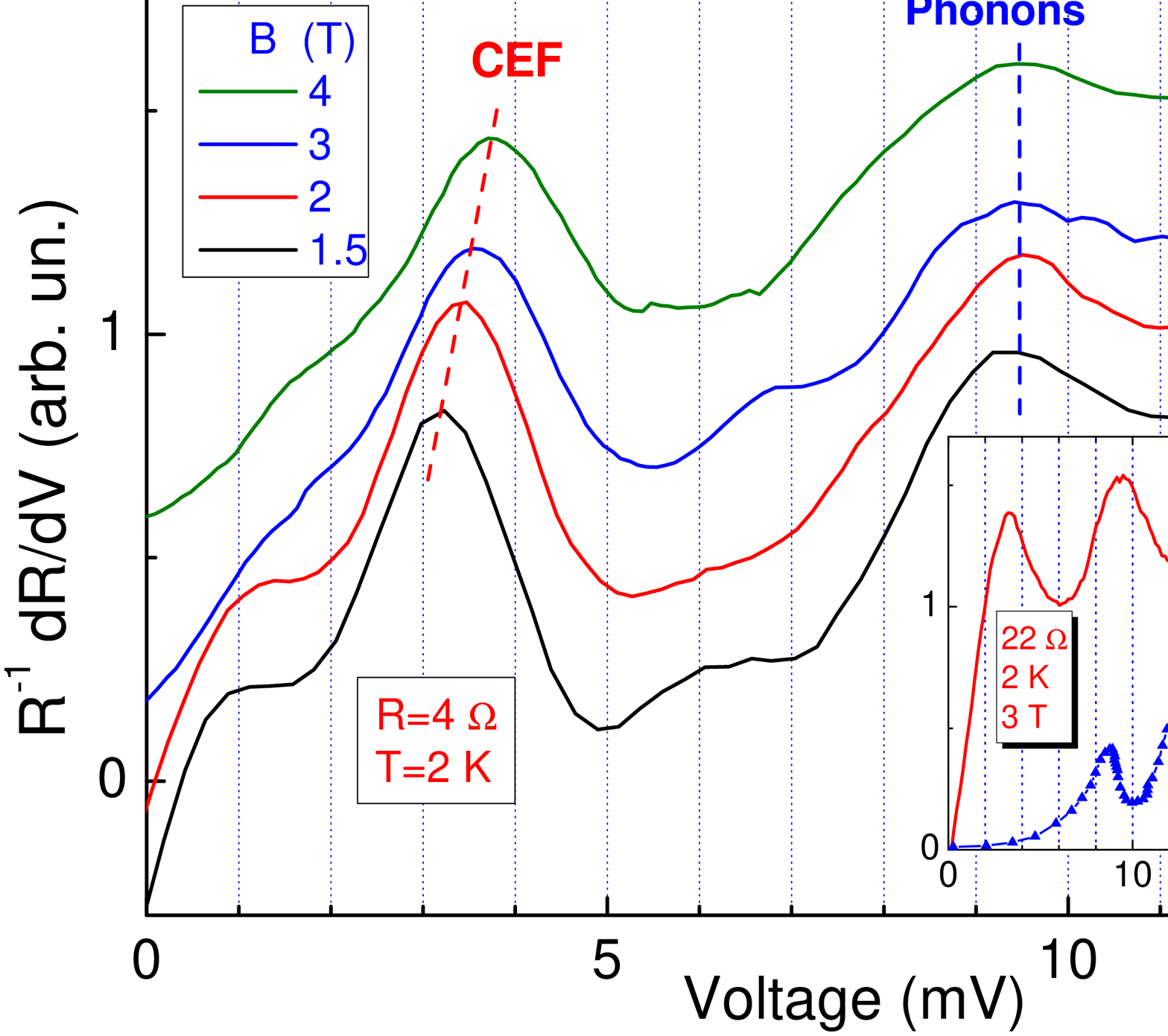}
\end{center}
\vspace{-3cm} \caption[] { (Color online) $R^{-1}dR/dV\propto
d^2V/dI^2$ curves measured at 2\,K in a magnetic field for the 3.8\,$\Omega$
PC from Fig.\,\ref{field}(b). The inset shows the PC spectrum of another
22\,$\Omega$ PC with sharper phonon maxima along with the phonon
DOS for LuNi$_2$B$_2$C (symbols). }
\label{d2vdi2}
\end{figure}
Now we consider the measured PC electron bosonic interaction spectra
($R^{-1}dR/dV = R^{-2}d^2V/dI^2$ curves) of TmNi$_2$B$_2$C--Cu
PCs. These spectra show clear maxima at about 3\,mV and 9.5\,mV
(see Fig.\,\ref{d2vdi2}) and a more smeared maximum at 15\,mV (see
inset in Fig.\,\ref{d2vdi2} and the following figures). The maxima at about
9.5\,mV and 15\,mV were not affected by magnetic field and
correspond well to the low-energy maxima in the neutron phonon
density of states of the nonmagnetic isostructural compound
LuNi$_2$B$_2$C \cite{Gompf} (see Fig.\,\ref{d2vdi2}, inset), only
the high-energy part of the obtained PC spectra remains almost
featureless. The displayed in the PC spectra
phonon features are similar to those observed in other nickel
borocarbide compounds. \cite{Bashlakov,naidyuk7,Naid07,Yanson97,YansonDy,Kvit10}
Thus, the observed peaks reflect the TmNi$_2$B$_2$C phonons excited by the inelastic
scattering of electrons (energized by e$V$-bias).
Unfortunately, even for the PC with a relatively large resistance (see
Fig.\,\ref{d2vdi2}, inset) for which the conditions for the
ballistic (spectroscopic) regime is fulfilled with a larger
probability, so far we could not resolve the phonon features above
20\,mV, contrary to the case of HoNi$_2$B$_2$C.\cite{Naid07} One
of the reasons might be that the PC spectrum of TmNi$_2$B$_2$C
contains a dominant 3-mV peak, which is connected with CEF
excitations as is shown below.
\begin{figure} [tbp]
\begin{center}
\includegraphics[width=7cm,angle=0]{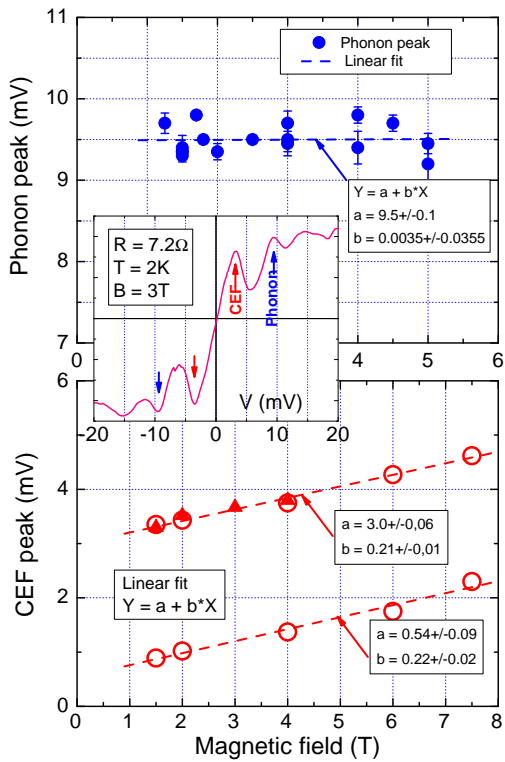}
\end{center}
\caption[] {(Color online) Upper panel: Position of the phonon
peak on $R^{-1}dR/dV\propto d^2V/dI^2$ PC spectrum (shown in
inset) for several TmNi$_2$B$_2$C--Cu contacts versus magnetic
field. Bottom panel: position of CEF peak(s) from
Fig.\,\ref{d2vdi2} (triangles) and Fig.\,\ref{d2vdi2b} (circles).
Dashed lines demonstrate linear fit of the data. Inset shows the
typical PC spectrum with mentioned peaks marked by the arrows.}
\label{d2vdi2a}
\end{figure}
Thus, excitations of Tm ions to the first CEF
level by e$V$-energized electrons followed by
subsequent scattering on 9.5-mV and 15-mV phonons will lead to a
shortening of the inelastic electron mean free path and to
deviations from the ballistic (spectroscopic) regime in our PC
with further increasing of bias e$V$.
This blocks energy-resolved spectroscopy at higher energies.

Next we will consider the mentioned low-energy peak near 3\,mV,
where the peak positions are marked by the tilted dashed line in
Fig.\,\ref{d2vdi2}. It can be assigned with the first
excited CEF transition in TmNi$_2$B$_2$C according to Ref.\cite{Gasser}.
As shown in Fig.\,\ref{d2vdi2a}, a magnetic field
shifts this maximum to higher energies whereas the maximum at
about 9.5\,mV remains fixed. This confirms phononic origin
of the 9.5-mV maximum. The other CEF transitions
at the higher energies mentioned in Ref.\cite{Gasser} could not be
resolved in the PC spectra, probably due to their lower intensity
and the mentioned deviation from the spectroscopic regime
with the bias increase. Thus, this low-lying first CEF excitation
seems to play a dominant role in the transport properties of
TmNi$_2$B$_2$C. The presence of such low-energy excitations in TmNi$_2$B$_2$C
can be anticipated from the nonsaturated resistivity in this compound
by lowering temperature with pronounced slope of $\rho(T)$ until
$T_{\rm c}$ (see Fig.\,\ref{field}(b), inset).

\begin{figure*} [htbp]
\vspace{4cm}
\begin{center}
\includegraphics[width=0.75\linewidth]{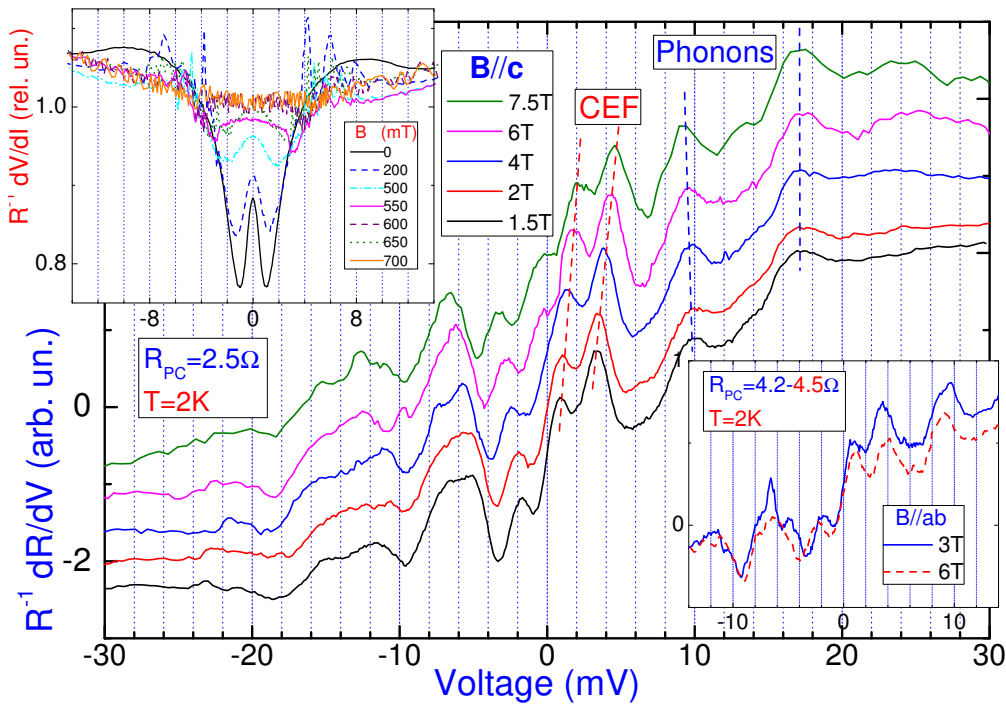}
\end{center}
\vspace{-4cm} \caption[] { (Color online) PC spectrum
$R^{-1}dR/dV$ of TmNi$_2$B$_2$C--Cu PC for two
bias polarity with additional 1\,mV peak vs magnetic field.
Left inset: $dV/dI$ for the PC from the main panel at low
fields, which also demonstrates "reentrant"
features above 500\,mT. Right inset: PC spectrum for another
PC at two fields showing a similar low-energy 1-mV peak
structure.} \label{d2vdi2b}
\end{figure*}

We have also occasionally measured  PC spectra with an additional
peak at low energy near 1\,mV (see Fig.\,\ref{d2vdi2b}). A
magnetic field shifts this peak as well as the
second peak at about 3\,mV (see Fig.\,\ref{d2vdi2a}, bottom panel)
to the higher energies, whereas the positions of the phonon maxima
near 10 and 17\,mV are stable. The 3-mV peak looks
like the CEF peak on the previous spectra. The 1-mV peak persists
up to the highest fields available in our experiment (7.5\,T)
exceeding the $H_{c2}$ in the $ab$ and $c$ directions
by a factor of four or two, respectively. Such a behavior
in a magnetic field excludes any residual
superconductivity in our PCs as the origin of this peak
\footnote{On the contrary, the hump near 1\,mV in  $dV/dI$ at 1.5 and 2\,T
in Fig.\,\ref{d2vdi2} is likely caused by residual
superconductivity because this structure disappears at higher
fields.}. The appearance of this 1\,mV peak
may testify to the nonstoichiometry of TmNi$_2$B$_2$C in the PC core.
In contrast to $R$=Ho, Er compounds, here in the region
of C and/or B vacancies or disordered C and B sites the original magnetic
Tm-ion ground state doublet splits into two singlet states
(see Ref.\cite{Gupta06}, page\,768). If so, then on
the one hand the PC spectroscopy enables us to distinguish the
"fine structures" in the CEF schema of TmNi$_2$B$_2$C, while
on the other hand it allows us to control the local stoichiometry.
Another possibility is that TmNi$_2$B$_2$C  exhibits a
quadrupolar ordered phase below $T_Q$=13.5\,K \cite{Andersen}.
The consequence of this would be that the quadrupolar ordering
lowers the CEF symmetry, resulting in the mentioned ground-state splitting.
In this context we note that a weak peak at about of 1\,meV
has been observed in the inelastic neutron spectra of
Tm$_{0.05}$Y$_{0.95}$Ni$_2$B$_2$C.\cite{Rotter} It was ascribed
by the authors to a transition derived from a splitted, originally
two-fold degenerate Tm-ion ground-state. Thus, we also assume here a
similar transition resulting from a modified CEF Tm-ion ground state,
either due to nonstoichiometry (vacancies) or due
to a quadrupolar ordered phase. To  finally elucidate the nature of
these peaks, more detailed directional measurements in higher
magnetic field with samples of well-defined stoichiometry are required as well.

\vspace{0.3cm}

\section{Conclusion}
We have carried out investigations of the SC gap
and of the electron-bosonic interaction spectra in TmNi$_2$B$_2$C
by means of AR and PC spectroscopy,
respectively. The SC gap in TmNi$_2$B$_2$C is found to deviate
considerably from the usual BCS-type behavior in showing a broad
maximum at $T^*\simeq T_{\rm c}/2$. An improved two-band analysis
of our data also shows the presence of a second SC gap being  about two times larger
than the former, which remains close to the small "first" gap observed.
The smaller gap has been assigned to a FSS
which has a similar orbital structure and a gap value as the
"cushion" FSS found in other
nickel-based 1221 borocarbide compounds. This not-yet-resolved FSS
for  TmNi$_2$B$_2$C is almost protected from
the exchange interaction with the magnetic moments of the
rare-earth ions.
The observed second larger gap is only a bit smaller
than the larger gap in nonmagnetic borocarbide superconductors.
 This way TmNi$_2$B$_2$C interpolates between
"cushion"-FSS-dominated almost single-band superconductors
DyNi$_2$B$_2$C and HoNi$_2$B$_2$C within the commensurate AFM state
and the nonmagnetic borocarbide superconductors YNi$_2$B$_2$C,
LuNi$_2$B$_2$C, and probably, also  the
ScNi$_2$B$_2$C compound which, however, is less well studied.

We succeeded to measure PC electron-bosonic interaction spectra of
TmNi$_2$B$_2$C with distinct phonon features at 9.5\,mV and
15\,meV along with intense low-energy maxima of
a nonphonon nature at about 3\,meV and additionally at about 1\,meV for some spectra.
The 3-meV maximum is due to excitation of Tm ions on the upper
(first) CEF level by e$V$-energized electrons. As to
the 1-meV peak, it is likely connected with a modification of the CEF
due to boron or carbon vacancies. This allows, in principle, for
the PC spectroscopy to be regarded as a local probe to check the stoichiometry.
Intense peaks of CEF
excitations in PC spectra point to their significant
contribution to the spectral function of the electron-boson interaction.

\section*{Acknowledgements}
Two of us, Yu.\ G.\ N. and O.\ E.\ K., thank IFW-Dresden for
hospitality and the Alexander von Humboldt Foundation for support.
Support by Pakt f\"ur Forschung at the IFW-Dresden
and the Deutsche Forschung Gemeinschaft as well as
the National Academy of Sciences of the Ukraine are gratefully
acknowledged. Discussions with  H.\ Rosner, K.-H.\ M\"uller,
and M. Schneider are kindly acknowledged.

\section*{Appendix}

As a rule, the parameter $\Gamma$ is used to describe the
nonthermal smearing of the experimental $dV/dI$ curves, namely,
the broadening of the characteristic AR minima in $dV/dI$, while
the original physical meaning of $\Gamma$ in the Dynes model
\cite{Dynes78} as a finite lifetime of charge carriers due to
inelastic scattering is usually not under consideration. This is
because other effects contribute to the $dV/dI$ broadening,  e.\,g., as SC gaps
distribution \cite{Pratap,Bobrov05} (via nonhomogeneity, surface, multiband, etc.
effects) or gap anisotropy
(see also \S4.3.4 in Ref. \cite{Daghero}).

Including $\Gamma$ in the fit procedure leads not only to the
smearing of the whole calculated $dV/dI$ curve but also results in
a reduced intensity of the whole AR (minima) structure. In this
context we note, that for a high-quality fit of $dV/dI$ in full,
not only the shape of $dV/dI$ must be described, but also its
absolute intensity must be correct.
In the present paper we also address this point, namely, we
considered the, so-called, scaling factor $S$, which is included to
fit the intensity of calculated and experimental curves. For
instance, $S$=1 means that the calculated curve fits the measured
$dV/dI$ also in absolute values. In the case of the ordinary
superconductor with a large coherence length like Zn, \cite{Naid96}
we succeeded in performing such a fit practically for all PCs.
If $S<$1, the measured $dV/dI$ has a reduced
intensity due to suppression of the AR
signal or the opening of some non-AR channel in  the PC
conductivity. Thus, the authors of Ref.\,\cite{Bobrov08}
assumed that $S<$1 is realized when the Fermi surface is only
partially gapped in the SC state. Of course,  the more common explanation
of non-SC regions in the PC or multicontact
scenario, including PCs with suppressed superconductivity, cannot
be ruled out as well. Thus, in Ref.\,\cite{Miyoshi} a
contribution to the conductance from the normal vortex cores in
the superconductor was taken into account to describe a progressive
suppression of the AR features in an external magnetic field.

It turned out that even a seemingly nonphysical situation with $S>$1
can also take place. As a rule,
large $\Gamma$ values result in large $S>1$. A more reasonable
explanation would be the following: if the smearing of $dV/dI$ is
due to a variation (distribution) of SC gaps over the Fermi
surface, then the description of this smearing by including
$\Gamma$ results in a reduced calculated $dV/dI$ intensity and
finally leads to a scaling parameter $S>1$. Hence such a
situation with $S>$1 might point to a gap distribution or to a
multiband scenario.

We should also mention that a contribution to the $dV/dI$ minima
may come from the Maxwell resistance, \cite{Naid} which cannot be
ignored for materials with a high residual resistivity like,
e.\,g., the heavy-fermion compounds. \cite{Gloos} Since the
residual resistivity in $R$Ni$_2$B$_2$C compounds is above
1$\mu\Omega$cm, then the Maxwell contribution to $R_{\rm {PC}}$
can exceed 1$\Omega$ for a PC with a diameter of 10\,nm. The
vanishing of the Maxwell resistance in the SC
state can lead to a situation where the PC resistance (the
conductance) can decrease (increase) more than 2 times, thus it
exceeds the limit of the maximal AR signal.


\end{document}